# Pre-metric electromagnetism as a path to unification.

DAVID DELPHENICH

*Independent researcher*
*Spring Valley, OH 45370, USA*
*feedback@neo-classical-physics.info*

It is shown that the pre-metric approach to Maxwell's equations provides an alternative to the traditional Einstein-Maxwell unification problem, namely, that electromagnetism and gravitation are unified in a different way that makes the gravitational field a consequence of the electromagnetic constitutive properties of spacetime, by way of the dispersion law for the propagation of electromagnetic waves.

**1. The Einstein-Maxwell unification problem.**

Ever since Einstein succeeded in accounting for the presence of gravitation in the universe by showing how it was a natural consequence of the curvature of the Levi-Civita connection that one derived from the Lorentzian metric on the spacetime manifold, he naturally wondered if the other fundamental force of nature that was known at the time – namely, electromagnetism – could also be explained in a similar way. Since the best-accepted theory of electromagnetism at the time (as well as the best-accepted "classical" theory to this day) was Maxwell's theory, that gave rise to what one might called the *Einstein-Maxwell unification problem:* Find some intrinsic (presumably geometric) structure on spacetime (suitably extended) that will decompose into (or at least lead to) the Lorentzian metric tensor $g$ and the electromagnetic field strength 2-form $F$, along with a set of field equations for that extended geometric structure that would imply the Einstein field equations for $g$ and the Maxwell equations for $F$ (at least, in some approximation).

The suspicion that such a "unified field theory" might actually exist was perhaps based upon the fact that Maxwell's theory itself represented a "unified field theory" of the electric and magnetic fields, and that there existed a well-known analogy between Coulomb's law of electrostatics and Newton's law of universal gravitation, although there were fundamental distinctions between them, as well. In particular, the analogy between mass and charge was not complete, since at the time (and to this point in time, as well), no one had ever observed what one might call "negative" mass or "anti-gravitation." Of course, the possibility that such a unification of gravitation and electromagnetism might lead to such tantalizing consequences has been an ongoing source of impetus for the search for that theory.

Several attempts followed by Einstein and others (cf., e.g., [1] and part II of [2]) at achieving such a unification. They seemed to fall into two basic categories: Extensions of $g$ with a four-dimensional spacetime and extensions of spacetime to something higher-dimensional. The former models included teleparallelism [1] and the Einstein-Schrödinger theory [2], while the latter include the Kaluza-Klein models [1, 2, 4, 5], and some of the attempts to extend the tangent bundle of the four-dimensional spacetime manifold to an anholonomic (i.e., non-integrable) rank-4 sub-bundle of the tangent bundle to a five-dimensional manifold (cf., e.g., [6-8]).

All of the attempts were regarded as failures for one reason or another. One problem with teleparallelism was that it included unphysical solutions, such as a static distribution of gravitating, uncharged masses. The main problem with the other theories was that they implied no new consequences of that unification; i.e., they were "concatenations" of the field theories, not unifications. What was missing were any sort of "gravito-electromagnetic inductions," that would suggest an analogue to the electromagnetic induction in Maxwell's theory. Indeed, it is important to note that the latter inductions had been established experimentally by Faraday before Maxwell formulated his theory, while to date, no such couplings of electromagnetic and gravitational fields seem to have materialized in the laboratory. ("Gravitomagnetism" is a different matter, and we shall discuss it below.)

---

[1] The author has compiled an anthology [3] of English translations of many of the early papers on teleparallelism that is available as a free PDF download at his website (neo-classical-physics.info)

However, as quantum physics evolved, the nature of the Einstein-Maxwell unification problem changed, as well. Increasingly, Einstein suspected that gravitation could only be unified with electromagnetism when one went to quantum electromagnetism. That possibility seems more reasonable nowadays, since the phenomenon of "gravitomagnetism" had not been observed experimentally until relatively recently [9]. The essence of that phenomenon is that the analogy between Coulomb's law and Newton's law goes beyond the scope of statics, since there is, in fact, a field that is induced by the relative motion of a mass that is analogous to the magnetic field that is induced by the relative motion of an electric charge, and which is commonly called the *gravitomagnetic field.* As a result, one sees that Maxwell's equations are closely analogous to the weak-field equations of gravitation, which suggests that perhaps Einstein's equations of gravitation, which are the strong-field equations, should be somehow analogous to some hitherto-unknown "strong-field" equations of electromagnetism.

The physical realm in which one would expect to find the strongest electromagnetic fields is in the atomic to subatomic domain, where one approaches the Schwinger critical field strengths at which photons resolve into electron-positron field-pairs. However, since the time of Heisenberg and Pauli [10], quantum electrodynamics has not started with a set of "strong-field equations of electromagnetism" that might perhaps be analogous to Einstein's equations of gravitation. Rather, it has simply passed over that "classical" problem [1] and started with the exchange-particle concept, combined with the scattering approximation for the field dynamics. That means: Rather than speculate on what might constitute the "field equations of QED" or the nature of the electromagnetic "force" that acts between elementary charges at the quantum level, quantum physics was going to replace the force of interaction with the exchange of an elementary particle that would mediate the interaction; for QED, that particle would be the photon. Then, rather than posing "classical" problems, such as boundary-value problems in statics and the Cauchy problem in dynamics, QED would simply pass to the approximation in which the initial time was − ∞ and the final time was + ∞, which is equivalent to assuming that the interaction of particles takes place inside a very small "black box" time interval in which the nonlinear nature of the interaction can be enclosed in such a way that the time evolution operator that takes incoming fields to outgoing ones becomes a linear operator that takes incoming scattering states (which are asymptotic free fields) to outgoing ones. That allows one to use the methods of Fourier analysis and discuss the scattering operator in momentum space without having to worry that the perturbation series that one defines (i.e., Feynman diagrams or loop expansions) is unphysical.

Of course, it is precisely the fundamental distinction between Einstein's theory of gravitation as a "classical" field theory (i.e., one in which one can pose boundary-value problems in statics and the Cauchy problem in dynamics), while QED is a "quantum" field theory (i.e., one that begins in the scattering approximation to that Cauchy problem) that is the greatest obstruction to the unification of those theories, although that fact is rarely addressed in quantum gravity, which takes more of a "play it where it lays" approach.

Another common critique of the Einstein-Maxwell unification problem is that it is currently a partial unification problem, in the sense that since the time of Einstein's early work on gravitation, two other "quantum" interactions have been added to the fundamental interactions, namely, the weak and strong interactions. Furthermore, once Yang and Mills had revisited the gauge-field approach to elementary interactions that Weyl, Fock, and Ivanenko had studied in the context of electromagnetism at about the same time that Einstein was pondering unified field theory and the Copenhagen school was defining their foundations for quantum physics, the unification of electromagnetism with the weak interactions as gauge field theories defined an entirely different approach to unification that usually took the form of looking for higher-dimensional gauge groups that might contain the more elementary gauge groups as subgroups. Interestingly, although gravitation was the first of the fundamental interactions to present a manifestly geometric character, and gauge fields also have a manifestly geometric character (as connection 1-forms), nonetheless, finding a gauge theory of gravitation that might be absorbed into the other gauge field theories in a unified way has proved to be more problematic than one would expect.

---

[1] One might suspect that the quantum use of the word "classical" in a pejorative way is probably an imitation of the pure-mathematical usage of the word "trivial," which often represents little more than a lack of personal curiosity about the subject, combined with an acceptance of the fact that the problem in question is hard to pose and even harder to solve.

## 2. Pre-metric electromagnetism.

Let us now consider another possibility, namely, that the Einstein-Maxwell unification problem is the wrong problem to pose. The justification for that is found in the fact that when one goes back to the chronological sequence of Einstein's early papers on relativity, one can notice a subtlety that is easy to ignore: Einstein did not start out looking for a theory of gravitation, he started out by examining the way that electromagnetic waves propagate from relatively moving bodies. It was the suggestion that the light-cones (i.e., characteristic manifolds for the propagation of those waves) represented the relativistically-invariant objects, when combined with the insight of Minkowski that the form of that characteristic equation for electromagnetic wave propagation suggested a non-Euclidian geometry on a four-dimensional space, that led Einstein to investigate other non-Euclidian geometries. In particular, Marcel Grossmann told him about Riemannian geometry (although light-cones are actually indicative of *pseudo*-Riemannian geometry), which eventually led to Einstein's theory of gravitation.

### 2.1 *The metric form of Maxwell's equations*.

Now, this chronological progression from electromagnetism to light-cones to gravitation makes perfect sense in the context of the "pre-metric" approach to electromagnetism. That approach is based upon the observation that the only place in which the Lorentzian metric on spacetime enters into Maxwell's equations is in the Hodge * operator. In order to see that, we express those equations in terms of the Minkowski electromagnetic field-strength 2-form $F$ as [1]:

$$dF = 0, \quad \delta F = 4\pi J, \quad \delta J = 0, \qquad (1)$$

in which $d$ represents the exterior derivative operator and:

$$\delta = \pm *d* \qquad (2)$$

represents the codifferential operator (whose sign will be negative for 2-forms on a four-dimensional Lorentzian manifold).

Now, let us express that * operator as the composition $\# \cdot C$ of two invertible linear maps. Namely, $C: \Lambda^2 \to \Lambda_2$ is the map that "raises both indices" of the 2-form $F$, so locally, one can express $C$ in components as:

$$C^{\kappa\lambda\mu\nu} = \tfrac{1}{2}(g^{\kappa\mu}g^{\lambda\nu} - g^{\kappa\nu}g^{\lambda\mu}). \qquad (3)$$

The other map is the *Poincaré isomorphism* $\# : \Lambda_2 \to \Lambda^2$, which is based upon a Riemannian volume element on spacetime, which takes the local component form:

$$V = \sqrt{-g}\ dx^0 \wedge dx^1 \wedge dx^2 \wedge dx^3, \qquad (4)$$

in which $g$ is the determinant of the component matrix $g_{\mu\nu}$ of the metric tensor.

The Poincaré isomorphism will then take the bivector field **B** to the 2-form #**B**, whose local components are:

$$(\#B)_{\mu\nu} = \tfrac{1}{2}\sqrt{-g}\ \varepsilon_{\kappa\lambda\mu\nu}B^{\kappa\lambda}. \qquad (5)$$

The 1-form $J$ represents the electric current that serves as the source of the field $F$, and is usually given the "convective" form $\sigma v$, where $\sigma$ is the electric charge density, and $v$ is the covelocity 1-form that is metric-dual to the velocity vector field for the moving source charge distribution.

---

[1] For the basic facts of this approach to Maxwell's equations, one might confer [11-14].

## 2.2 The pre-metric form of Maxwell's equations.

The observation of Kottler [15], which was subsequently pointed out by Cartan [16], and expanded upon by van Dantzig [17] was that, in a sense, the linear isomorphism $C$ (which is where $g$ enters into Maxwell's equations) plays the role of an *electromagnetic constitutive law*. [18, 19]. Generally, such a law associates the *electromagnetic excitation* bivector field $\mathfrak{H}$, which includes the electric displacement $D$ and the magnetic flux density $H$, with the electromagnetic field strength $F$, which includes the $E$ and $B$ fields [1]. Of course, the more general association:

$$\mathfrak{H} = C(F) \qquad (6)$$

does not, by any means, have to be linear on the fibers of the bundles in question. Indeed, nonlinear electromagnetism seems to be an unavoidable aspect of the "strong-field" form of Maxwell's equations (whatever that might be). In fact, even the restriction to invertible maps on fibers represents the restriction to "non-dispersive" media, which would make $C$ and algebraic operator, instead of an integral one.

In the event that linearity is an acceptable approximation, the map $C$ can be represented by a fourth-rank tensor field whose local components take the form $C^{\kappa\lambda\mu\nu}(x)$, such that:

$$\mathfrak{H}^{\mu\nu} = \tfrac{1}{2} C^{\kappa\lambda\mu\nu} F_{\kappa\lambda} . \qquad (7)$$

The most elementary constitutive law that one can impose upon a medium is precisely the one that one implicitly uses in special relativity. If one assumes that the spacetime vacuum is characterized by being non-dispersive, linear, isotropic, and homogeneous then one can use:

$$D^i = \varepsilon_0 E^i, \qquad B^i = \mu_0 H^i, \qquad (8)$$

in which $\varepsilon_0$ is the classical vacuum dielectric constant and $\mu_0$ is its magnetic permittivity. Note that one still needs a spatial metric in order to raise the indices on $E$ and $H$.

The fact that one rarely considers the electromagnetic constitutive properties of spacetime in either special or general relativity is largely due to the fact that one usually encounters them only in the combination:

$$c_0 = \frac{1}{\sqrt{\varepsilon_0 \mu_0}} , \qquad (9)$$

which is then set equal to 1.

One then sees that the map (3) of raising both indices of a 2-form does, in fact, take the form of a special case of an electromagnetic constitutive law [2].

If one replaces the Riemannian volume element with the more general one:

$$V = dx^0 \wedge dx^1 \wedge dx^2 \wedge dx^3, \qquad (10)$$

so:

$$(\#B)_{\mu\nu} = \tfrac{1}{2} \varepsilon_{\kappa\lambda\mu\nu} B^{\kappa\lambda}, \qquad (11)$$

this time, then one can express Maxwell's equations in their "pre-metric" form [3]:

$$dF = 0, \ \operatorname{div} \mathfrak{H} = 4\pi \mathbf{J}, \ \operatorname{div} \mathbf{J} = 0, \ \mathfrak{H} = C(F), \qquad (12)$$

in which:

$$\operatorname{div} = \#^{-1} d \# \qquad (13)$$

is the adjoint of $d$ (and which agrees with the usual divergence operator on vector fields), and $\mathbf{J} = \sigma \mathbf{v}$ is the electric current vector field.

One can also absorb the map $C$ into the basic equations and arrive at:

$$dF = 0, \ \operatorname{div} C(F) = 4\pi \mathbf{J}, \ \operatorname{div} \mathbf{J} = 0. \qquad (14)$$

Furthermore, if one chooses an electromagnetic potential 1-form $A$ (so $F = dA$) then this will reduce to:

$$\Box_C A = 4\pi \mathbf{J}, \qquad \operatorname{div} \mathbf{J} = 0, \qquad (15)$$

in which we have introduced the generalized d'Alembertian operator that is associated with the map $C$:

---

[1] Actually, this association becomes confused somewhat by the fact that in the electrodynamics of continuous media [19], the **B** field is the response of the medium to the imposition of the **H** field, not the other way around.

[2] The author has recently investigated the electromagnetic interpretation of some of the most popular classes of Lorentzian metrics in [20].

[3] This form of the pre-metric equations is due to the author [13]. A different, but equivalent, form is found in Hehl and Obukhov [14].

$$\square_c = \text{div} \cdot C \cdot d. \quad (16)$$

## 2.3 *The emergence of light cones.*

In order to see how one gets back to a spacetime Lorentzian structure, one first reminds oneself that the reason that one says "light-cones" instead of "gravity-cones" is that the two theories are not independent of each other: The light-cones are characteristic hypersurfaces for the propagation of electromagnetic waves, and they also represent of the dispersion law for those waves. Those light-cones also relate to the fundamental structure that implies the presence of gravitation in spacetime, namely, the Lorentzian metric.

In fact, when one goes to pre-metric electromagnetism, one finds that the quadratic form of the light-cone equation:

$$g(\mathbf{v},\mathbf{v}) = \eta_{\mu\nu} v^\mu v^\nu$$
$$= c^{-2}(v^0)^2 - (v^1)^2 - (v^2)^2 - (v^3)^2 \quad (17)$$

is a degenerate case of a more general quartic expression.

In order to get that quartic expression, one first needs to restrict the scope of the theory to linear electromagnetic media, since otherwise, one would have to expect that the dispersion law for "wave-like" solutions of the field equations (14) would also depend upon the definition of "wave-like." For linear media, it is entirely sufficient to find the dispersion law for plane-wave solutions, even though they have a distinctly unphysical character, due to their infinite total energy and momentum. For such fields, one can locally set:

$$F = e^{-ik(\mathbf{x})} f, \quad (18)$$

in which:

$$k = \omega\, dt - k_i\, dx^i \quad (19)$$

is the *frequency-wave number 1-form* for a wave and:

$$\mathbf{x} = x^\mu \frac{\partial}{\partial x^\mu} \quad (20)$$

is the *position vector field* that is defined by the choice of coordinate chart. The field $f$ basically represents the shape of the wave; in the geometrical optics approximation, one effectively sets $df = 0$.

If one substitutes (18) into (15), and considers only the points of spacetime that are outside the support of **J** (so **J** = 0) then after some tedious, but straightforward, calculations [1], one will get a linear map $L(k): \Lambda^1 \to \Lambda_1$ that is quadratic in $k$. It is not invertible, since the first step in the composition of maps that gives one $L(k)$ takes $A$ to $k \wedge A$, which will be zero for any $A$ that are collinear with $k$, so one must first restrict $L(k)$ to a complementary subspace to the line that is generated by $k$. In fact, in order to get an invertible map, one must reduce to a two-dimensional subspace of that three-dimensional subspace, and if $L_2(k)$ is the restriction of $L(k)$ to that two-dimensional subspace then the condition for the invertibility of $L_2(k)$ is the non-vanishing of the determinant of that linear map.

That determinant will, of course, depend upon $k$, and the *characteristic* $k$ are the one for which the determinant vanishes:

$$D_4(k) \equiv \det L_2(k) = 0. \quad (21)$$

The subscript 4 indicates that the function $D_4$ is a homogeneous polynomial of degree 4 in $k$; in fact, it will generally be quadratic in $k^2$. Equation (21) then represents the characteristic hypersurfaces for the electromagnetic waves that propagate in spacetime according to (12), (14), or (15), as well as the dispersion law for such waves.

As a polynomial of degree four, the function $D_4$ can also be associated with a completely-symmetric, covariant, fourth-rank tensor field on spacetime that has been called the *Tamm-Rubilar tensor* [14]. Its form is very closely related to the study of *Kummer surfaces* [21], which grew out of the branch of projective geometry that is called *line geometry* [22].

The general electromagnetic medium will exhibit *birefringence*, which means that if one first treats the components $k_\mu$ as the homogeneous coordinates for a point in $\mathbb{R}P^{3*}$ then the corresponding inhomogeneous (i.e., Plücker) coordinates $n_i = k_i / \omega$, $i = 1, 2, 3$, will take the form of indices of refraction in the three elementary directions of space. If one then represents $n_i$ as $n\, u_i$, where $u_i$ are the components of a unit vector in the (spatial) direction of propagation, then the equation $D_4(n_i) = 0$ will generally have two distinct roots for $n^2$; i.e., the same direction of propagation will be associated with two different speeds of propagation.

---

[1] For more details on the calculations, one can confer the books by the author [13] and Hehl and Obukhov [14].

(Although it is not clear in the present context, the distinction relates to the state of polarization of the wave.)

A first reduction in generality for $D_4$ is to the product of quadratic functions:

$$D_4(k) = D_2(k) D_2'(k), \qquad (22)$$

in which $D_2(k)$ and $D_2'(k)$ are homogeneous, quadratic polynomials in $k$, and generally of Lorentzian type. One calls this possibility *bi-metricity* [23], and it basically represents a pair of distinct light-cones at every point.

The final reduction that brings one back to Lorentzian structures of the kind that are treated in general relativity is to look at only constitutive laws for which:

$$D_2(k) = D_2'(k) = g(k, k), \qquad (23)$$

so:

$$D_4(k) = g(k, k)^2. \qquad (24)$$

In particular, such a medium cannot be birefringent.

## 3. Conclusion

One can now see how many restricting assumptions must go into starting from the electromagnetic constitutive properties of the spacetime manifold and the pre-metric field equations of the electromagnetic field strength 2-form $F$ and concluding with a Lorentzian metric $g$. In particular, one must assume that the medium is non-dispersive, linear, and non-birefringent. In fact, many of the popular forms that $g$ takes in general relativity also prove to be spatially isotropic. In effect, the only room for variety in the gravitational field is when the medium is not electromagnetically homogeneous.

It is the restriction to linear media that defines the strictest limitation in the eyes of quantum electrodynamics, since one generally finds that the effective electromagnetic field equations (such as Heisenberg-Euler, which are one-loop effective equations) are nonlinear generalizations of Maxwell's equations that also involve nonlinear effective constitutive laws. In fact, they exhibit what is commonly called *vacuum birefringence*; that is, the vacuum polarization that is associated with the electromagnetic field at the Schwinger point also breaks down the light cone structure into a bimetric structure.

The conclusion that we have been leading up to through all of this is that although pre-metric electromagnetism does not provide a solution to the Einstein-Maxwell unification problem, it does nonetheless exhibit a radically different approach to the unification of the two field theories, which is that *the gravitational field in spacetime comes about as a consequence of the electromagnetic constitutive properties of spacetime*. Hence, the field $C$ is more fundamental to spacetime structure than $g$; in effect, gravitation is the shadow that is cast by electromagnetism.

Something that was only touched upon here that is also a radical departure from the usual approach to the geometry of spacetime is that line geometry is to electromagnetism what metric geometry is to gravitation. Hence, the very type of geometry that one is considering changes, as well. (For more on that aspect of the problem, see the author's work [13, 22].)

## References [1]

---

[1] The references that are marked with an asterisk are available in English translation at the author's website (neo-classical-physics.info)